# A template method for carbon nanotube production from sugar water

V.Y. Butko[a,b,*], A.V. Fokin[a], V.N. Nevedomsky[a], and Y.A. Kumzerov[a]

[a]Ioffe Physical Technical Institute, Russian Academy of Science (RAS), 26 Polytechnichesaya, St. Petersburg 194021, Russia.

[b]St. Petersburg Academic University, Nanotechnology Research and Education Centre, RAS, 8/3 Khlopin, St. Petersburg, 195220, Russia.

## Abstract

The methods of carbon nanotubes (CNTs) growth inside nanoporous materials are among the best candidates to fabricate CNTs with controlled geometrical parameters. We report obtaining CNTs by pyrolysis of sugar water and show that this chemical reaction is better suited for various applications than other known template methods. The diameter of the obtained CNTs, their alignment, and the distance between them can be controlled by choosing the bendable nanoporous template (chrysotile). Electron microscopy shows good crystalline quality of the obtained approximately 10 μm long CNTs after the template etching. Quasi-metallic transport is observed in this material after annealing CNTs under inert atmosphere.

---

[*] Corresponding author. Tel.: +78125980230; fax: +78125156747.
E-mail address: vladimirybutko@gmail.com (V.Y.Butko).

# 1. Introduction

Carbon nanotubes (CNTs) have excellent mechanical, thermal, and electronic properties for a variety of applications. To realize the full potential of CNTs an inexpensive method that is capable of obtaining them in large quantities in proper positions, with controlled diameter and correct alignment is needed. This problem has been recognized for the development of biomedical and other filters[1], catalytic systems [2], heating and cooling elements[3], receiving and transmitting antennas[3], terahertz polarizers [4], field emission devices[5,6], infrared absorbers[7], three dimensional holograms[8], supercapacitors[9,10], transistors[11] , and other electronic devices[12, 13 ]. To solve this problem one can chemically grow carbon nanotubes of a chosen diameters that are positioned, and aligned as needed for specific applications. The template methods of carbon nanostructure fabrication are among the best candidates to fulfill these goals. However, the known template methods are based on catalytic carbon deposition[12] or on pyrolysis of organic materials[14,15] and produce not only carbon nanostructures but leave remains of catalysts and other chemical ingredients. These ingredients can cause defects and complicate fabrication of long crystalline carbon nanotubes. Moreover the catalytic carbon deposition for CNT template fabrication inside nanoporous channels with the smallest diameters seems to be too difficult to achieve[16]. The nanoporous templates used for CNT fabrication so far [12,14] are also not well suited for the above goals due to rigidity of these templates.

# 2. Sample preparation

We have developed a method of production of CNTs inside bendable nanoporous templates filling them with syrup. Our method resembles candy fabrication from syrup heated inside lollipop molds. This production comprises preparation of the syrup by dissolving sugar (sucrose) in water, filling this syrup into a porous template, and the following high temperature annealing. We show that pyrolysis of sugar water during this annealing under inert atmosphere inside nanochannel porous templates produce long CNTs of high crystalline quality. An advantage of this chemical reaction for carbon nanotube fabrication compared to the chemical reactions that were previously used for this purpose[11-16] is related to the fact that pyrolysis of sugar water produces just carbon and water without any other chemical ingredients that can be harmful for the nanotube growth. This method also does not require any catalysts unlike the methods of carbon nanostructure catalytic growth [11,16,17]. Another advantage of this method is that a large number of nanoporous materials can potentially be used as templates because they can be easily filed by sugar water due to good adhesion of syrup to various surfaces. Therefore pyrolysis of sugar water is potentially an important for production of carbon nanotubes inside templates.

We demonstrate this method using nanochannels of chrysotile nanofibers as a template. Its idealized chemical formula is $Mg_3(Si_2O_5)(OH)_4$. This material is obtained from natural serpentinite rocks which are common throughout the world. The chrysotile template can be completely removed from the nanotube product at the final stage of the method. Properties of the nanoporous chrysotile nanofibers are quite favorable for use as templates. The individual nanofibers within the bundles (see Figure 1(a) and Figure 1(b)) have a tendency to be very long (up to ~10 cm). They have sufficient flexibility and tensile strength so they can be controllably aligned and woven into cloth. Similar nanofibers were used earlier as templates for fabrication of inorganic nanowires see for instance[18,19]. Here we use nanofibers with the inner diameter d of approximately 10 nm as a template for carbon nanotube growth. We have found that nanoporous channels in these nanofibers can be filled with a solution of sugar in water under normal atmospheric pressure. The nanoporous templates filled with this solution have been annealed in air in the temperature range 200÷500 ºC for approximately 24 hours. After that these samples were etched in HCl acid for approximately 48 hours at room temperature to remove Mg component of the nanofibers. This procedure was followed by the samples annealing at temperatures between 800 ºC and 1200 ºC in argon or in a high vacuum. The typical time of the annealing was approximately 1 hour.

Annealed samples became uniformly black. After the second annealing for investigation the template was completely removed by etching in HF acid or in HCl and HF 50% mixture for approximately 48 hours at room temperature.

## 3. Sample investigation

The fabricated samples have been studied by Transmission Electron Microscopy (TEM) and transport measurements. TEM measurements have been made after ultrasound treatment of the samples in ethanol by using JEM 2100F microscope with 200 kV accelerating voltage. The obtained TEM images are shown in Figures 2a, 2b and 2c. As one can see in these figures nanotubes have been obtained. The existence of unbroken twisted nanotubes seen in Figure 2a indicates robustness of their structure. Ordered crystalline structure of the obtained nanotubes with atomic interlayer distance of ~3.6 A can be seen in Figure 2(c). This value matches atomic interlayer distance reported in[20] (the interlayer distance in CNTs ranges from 0.342 to 0.375 nm). This match confirms the carbonous nature of the nanotubes obtained here by pyrolysis of sugar water. Some of these nanotubes are approximately 10 μm long. This length is longer than the length of graphene wall nanotubes fabricated by other pyrolytic template methods[14].This length was measured after the ultrasound treatment of the nanotubes. This treatment is considered to be a destructive procedure. Therefore even longer nanotubes are likely to be obtained without ultrasound treatment. Typical internal and external diameters of the obtain CNTs are 6 nm and 10 nm correspondently. This external diameter of the obtained CNTs matches the diameter of the porous nanochannels inside the used chrysotile fibers (d). This match confirms that the nanochannel pores in chrysotile play the role of the template for CNT growth. Therefore unbroken by ultrasound nanobundles that one can see in Figures 2(a) and 2(b) likely comprise aligned arrays of CNTs. These compelling evidences indicate the formation of nanotube arrays, but does not eliminate some degree of uncertainty about nanotube position and alignment after the template etching. However, Fig. 3 demonstrates that the initial alignment and the diameter of the bundle can be preserved to a great extent through this process. Therefore the proposed method allows the fabrication of controllably aligned carbon nanotubes with a determined distance between neighboring nanotubes in the perpendicular to the bundle main axis direction. As one can see from Fig.1(a) this distance is mainly determined by the external diameter of chrysotile fibers (D). D value depends on the type of chrysotile and is in the range from 15 nm to 40 nm.

Four-probe low temperature transport measurements in the obtained materials with silver paste contacts separated by a distance of ~1 mm have been performed in a liquid nitrogen dewar. Temperature was measured by using copper-constantan thermocouple. The four-probe resistivity versus temperature for two typical samples is shown in Fig. 4. The observed general trend is that the sample resistivity decreases with the increase of the annealing temperature. Disorder typically decreases conductivity in organic single Field Effect Transistors[21]. The same dependence was also established in an inorganic material due to Coulomb gap opening[22,23]. Therefore our observation is consistent with a conclusion that higher annealing temperature provides better crystalline quality of the obtained graphene nanostructures. As one can see in Fig.4 a quasi metallic behavior has been reached in the samples annealed at 1200 ºC. To estimate conductivity of the carbon nanostructures in this etched sample its porosity needs to be taken into account. This estimate is of the order of ~500 S cm$^{-1}$ at room temperature (this value is obtained from the data shown in Fig. 4 anda typical porosity of the chrysotile template estimated as ~7%). This conductivity is similar to the values measured in carbon nanotube ropes[24]. As one can see in Fig.4 the samples annealed at temperatures above 1080 ºC demonstrate similar or even more metallic temperature dependence of resistivity compared to the slightly semiconducting one that is typically observed in this temperature range in high quality CNT bundles[25]. The observed properties confirm a good crystalline quality of the obtained CNTs.

## 4. Conclusion

A new functionality of the obtained CNTs arises from the unique features of the proposed method. This functionality includes high degree of control over nanotube alignment, over diameter of the nanotubes, and distance between them that can be attained by choosing parameters of the nanoporous templates. It is conceivable that additional optimization of the nanotube growth would significantly increase length and decrease diameter of the obtained nanotubes. The upper limit to this length and the lower limit to this diameter are determined by the known maximal length and minimal internal diameter of the chrysotile nanochannels that are of the order of ~10 cm and ~2 nm correspondingly. Simplicity of the proposed method, low cost of the required materials, and demonstrated possibility to fabricate controllably curved crystalline conductive carbon nanotubes with chosen diameter make this method useful for developing superior quality catalytic devices with a larger internal surface area, nanofilters, polarizing electromagnetic devices, field emission cathodes, and other electronic devices.

## Acknowledgements

We are grateful to A.V. Babichev, A.A. Sisoeva, I.A. Izmailova, A.V. Butko, and A.Y.Egorov for help. This work was supported by the Russian Foundation for the Basic Research under project № 10-02-00853-a, and 11-02-0739-a, under Presidium of the RAS project, under the project PHANTASY and the Collaborative European Projects EU-RU.NET.

**Figure Captions**

**Figure 1(a)**

A cartoon of porous nanofibers used as a template (mold) for nanotube fabrication.

**Figure 1(b)**

A chrysotile bundle. It is lighted from the left by a laser beam. The absence of light scattering inside the bundle demonstrates high quality packing of nanoporous chrysotile.

**Figure 2 (a)**

A TEM image of typically obtained carbon nanotubes (1) and nanobundles (2).

**Figure 2 (b)**

A TEM image of typically obtained carbon nanotubes (1) and nanobundles (2).

**Figure 2 (c)**

A high resolution TEM image of the carbon nanotube region.

**Figure 3**

The obtained material comprising carbon nanotubes inside template after it has been annealed at 920 C in argon. This sample is on top of a millimeter scale.

**Figure 4**

Results of four-probe low temperature transport measurements in the obtained samples.

**Figures**

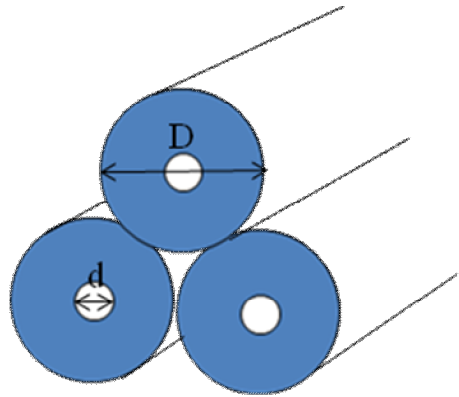

**Figure 1(a)**

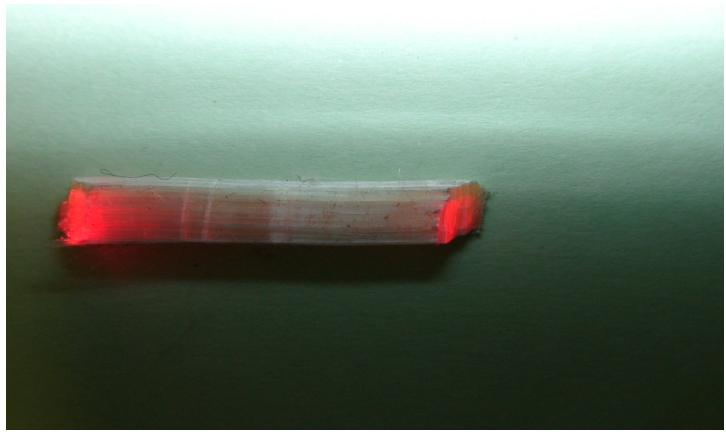

**Figure 1(b)**

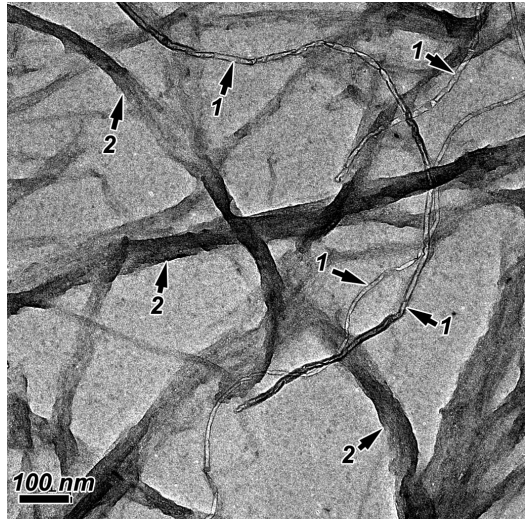

**Figure 2 (a)**

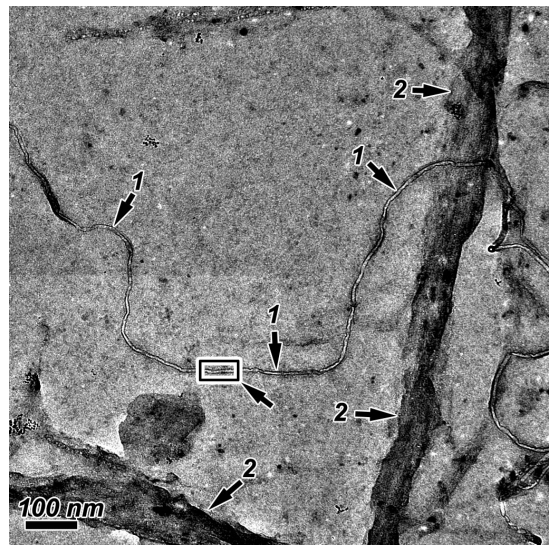

**Figure 2 (b)**

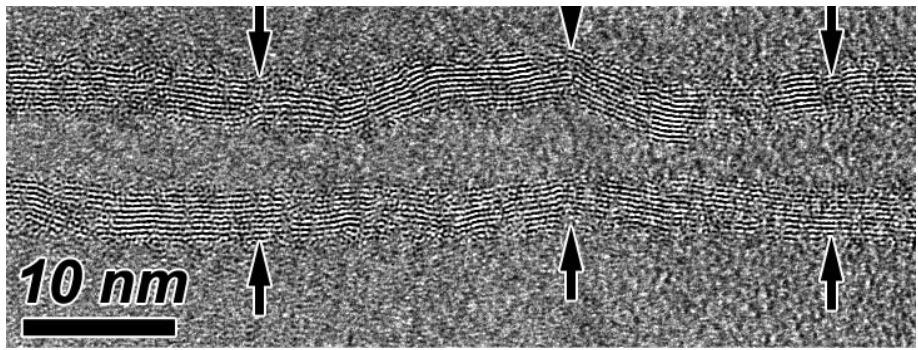

**Figure 2 (c)**

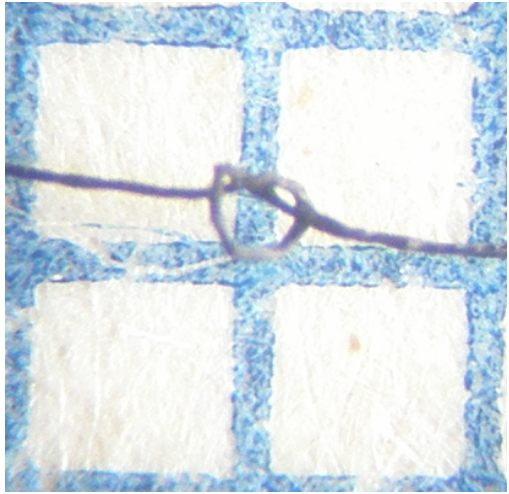

**Figure 3**

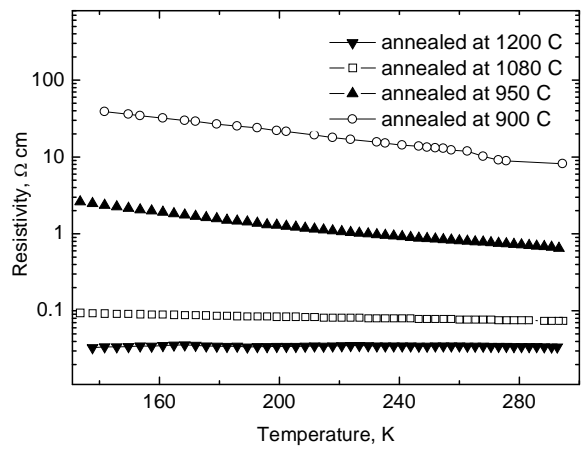

**Figure 4**